\documentclass{aa}
\usepackage{graphicx,psfig}

\def \fuse{{\it FUSE}}
\def \iue{{\it IUE}}
\def \ie{i.e.}
\def \eg{e.g.}
\def \g231{G$\,$231--40}

\begin{document}
   \title{Quasi-molecular lines in Lyman wings of cool DA white~dwarfs}
   \subtitle{Application to \fuse\ observations of G$\,$231--40}

   \author{    G. H\'ebrard    \inst{1}
          \and N.~F. Allard    \inst{1,2}   
          \and I. Hubeny       \inst{3} 
          \and S. Lacour       \inst{4,1}
          \and R. Ferlet       \inst{1} 
          \and A. Vidal-Madjar \inst{1}}

   \offprints{G. H\'ebrard (hebrard@iap.fr)}

   \institute{Institut d'Astrophysique de Paris, CNRS, 98$^{bis}$ boulevard 
              Arago, F-75014 Paris, France 
   \and  
 Observatoire de Paris-Meudon, LERMA, F-92195 Meudon Principal Cedex, France 
   \and
             NOAO, 950 North Cherry Avenue, Tucson, AZ 85726,  USA  
   \and 
             Department of Physics and Astronomy, Johns Hopkins University, 
             Baltimore, MD 21218, USA 
  }

   \date{Received ...; accepted ...}

   \abstract{
We present new theoretical calculations of the total line
profiles of  Lyman~$\alpha$ and Lyman~$\beta$  which include
perturbations by both neutral hydrogen {\em and} protons and all possible
quasi-molecular states of H$_2$ and H$_2^+$. They are used to improve
theoretical modeling of synthetic spectra for cool DA white dwarfs. 
We compare them with \fuse\ observation of \g231. 

The appearance of the
line wings between Lyman~$\alpha$ and  Lyman~$\beta$ is shown to be
sensitive to the relative abundance of hydrogen ions and neutral atoms,
and thereby to provide a temperature diagnostic for
stellar atmospheres and laboratory plasmas. 

   \keywords{Line: profiles -- Radiation mechanisms: general -- 
             Stars: atmospheres -- Stars: individual: G231-40 -- 
             White dwarfs -- Ultraviolet: stars}
   
}

   \maketitle
%

\section{Introduction}
Structures in the Lyman~$\alpha$ and Lyman~$\beta$ line wings have 
been identified with
free-free transitions which take place during binary close collisions
of the radiating H atom and a perturbing atom or ion 
(Allard~et al.~1998a, 1998b, 1999).
The characteristics of these features (position, amplitude, and
shape), due to the formation of
quasi-molecules during collisions between the radiating atom and
perturbers, depend directly on the potential energy curves correlated
to the atomic levels of the transition (Allard \& Kielkopf 1982).

Two satellite absorption features at 1058~\AA\ and 1076~\AA\ 
due to collisions of atomic hydrogen 
with protons were first identified in the  spectrum of the
DA white dwarf Wolf$\,$1346, as observed with the Hopkins Ultraviolet
Telescope (Koester~et al.~1996). 
These satellites in the red wing of Lyman~$\beta$ are in the 
Far Ultraviolet Spectroscopic Explorer (\fuse) spectral range 
(Moos et al.~2000); 
furthermore, Lyman~$\beta$ profiles are also the subject of an ongoing 
study of the far ultraviolet spectrum of dense hydrogen~plasmas.

In Allard~et al.~(1998a) we 
presented  theoretical profiles of Lyman~$\beta$ perturbed  solely by
protons. The calculations were based on the
accurate theoretical H$_2^+$ molecular potentials of
Madsen \& Peek~(1971) 
to describe the interaction between radiator and perturber,  
and dipole transition moments of
Ramaker \& Peek~(1972). 
The line profiles were included
as a source of opacity  in model atmospheres for hot white dwarfs, 
and the predicted spectra compared well with the observed 
{\it ORFEUS} and \fuse~spectra (Koester~et al.~1998; Wolff~et al.~2001).

{\it Ab initio} calculations of Drira~(1999) of electronic transition
moments for excited states of the H$_2$ molecule
and molecular potentials of  Detmer~et al.~(1998)
allowed us to compute Lyman~$\beta$ profiles perturbed
by neutral atomic hydrogen (Allard~et al.~2000).
The appearance of a broad  satellite situated at 1150~\AA\ makes 
 necessary to take into 
account the total contribution of both the  Lyman~$\alpha$ and
Lyman~$\beta$ wings of H perturbed simultaneously by neutrals and
protons.

We show  that the shape of the wings in
the region between Lyman~$\beta$ and Lyman~$\alpha$ is
particularly sensitive  
to the relative abundance of the neutral and ion perturbers responsible
for the broadening of the lines.

These new profiles  have been used to predict synthetic spectra for
cool DA white dwarfs which present  structures
at 1600~\AA\ and 1400~\AA\ in the Lyman~$\alpha$ wing
due respectively to quasi-molecular absorption of the H$_2$ and H$_2^+$
molecules. These last two structures have been demonstrated to 
be a very sensitive temperature indicators in DA white dwarfs.
The relative strength of these two satellite
features depends very strongly on the degree of ionization in the
stellar atmosphere, and thus on the stellar parameters $T_{\rm eff}$ and
$\log g$ (Koester \& Allard~1993; Koester~et al.~1994; 
Bergeron~et al.~1995).

\section{Theoretical line profiles}

\subsection{Theory}

We use a general unified theory in which the electric dipole moment
varies during a collision; a detailed description of the theory as
applied to the shape of the Lyman lines 
has been given by Allard~et al.~(1999). The obtained
line profiles fit the spectra of laser-produced 
hydrogen plasmas (Kielkopf \& Allard 1998).

 Our approach requires  prior knowledge of accurate theoretical molecular
potentials to describe the interaction between radiator and perturber, and  
knowledge of the variation of the radiative dipole moment with
atom-atom and atom-ion separation for each
molecular state. This effect is important when the dipole moment varies
in the region of inter-nuclear distance where the satellite is formed, 
and thus cannot be neglected. 
In the case of Lyman~$\beta$ satellites, due to 
H-H$^+$ collisions, we have shown 
that large changes (up to 60~\%) 
in the intensity of 
the satellites may occur when the variation of the dipole moment is
taken into account (Allard~et al.~1998a). This result is also valid
for other  lines; it increases by a factor of about 2
the main satellites of Lyman~$\alpha$ (Allard~et al.~1999).
Previous line profile calculations using constant dipole moment have
been  used  to  interpret \iue\ (International Ultraviolet Explorer) 
and Hubble Space Telescope
spectra by Koester \& Allard~(1993), Koester~et al.~(1994), and 
Bergeron~et al.~(1995). The synthetic spectra presented here used
improved  Lyman~$\alpha$ line profiles of Allard~et al.~(1998b)
which have been already included in stellar atmosphere
programs for the computation of  stellar atmosphere model
and synthetic spectra of $\lambda$~Bootis~stars. 
A comparison of these calculations with observations made with the \iue\ 
demonstrated that these last improvements are of fundamental
importance for  obtaining a better quantitative interpretation of the 
spectra and for determining stellar atmospheric parameters (Allard~et 
al.~1998b).

 \subsection { Lyman $\beta$ in H-H  collisions}

The Lyman profiles and satellites
are  calculated at the low densities met in the atmospheres of stars.
The typical particle densities (10$^{15}$ to 10$^{17}$ cm$^{-3}$) allows
us to  use an expansion of the autocorrelation function in powers of
density  as described in Allard~et al.~(1994) and Royer~(1971).
 Line profiles are 
normalized so that over $1/\lambda=\omega$ (cm$^{-1}$) they integrate to 1.

\begin{figure}      
\psfig{file=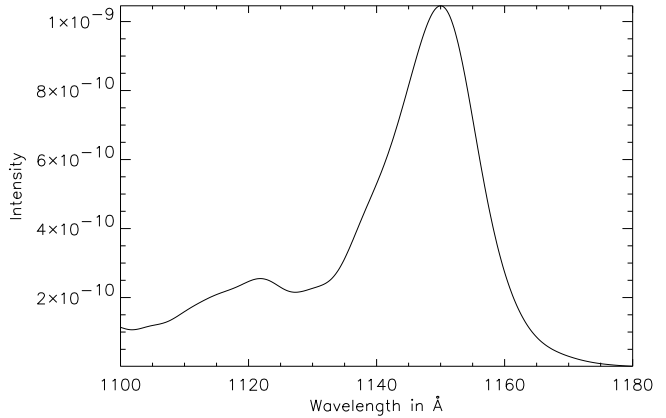,width=9.0cm}
\caption{Lyman~$\beta$ satellite due to H-H collisions. The
neutral density $n_{H}$ is $1\times10^{16}$ cm$^{-3}$.}
\label{beta}
\end{figure}

The only line feature of the  Lyman~$\beta$  profile is a broad 
absorption line satellite situated at 1150~\AA\ due to the
$\mathrm{B}''\bar{\mathrm{B}}\;^1\Sigma^+_u - \mathrm{X}\;^1\Sigma^+_g$ 
molecular transition of H$_2$ (Fig.~\ref{beta}).  
 The recent {\it ab initio} calculations of 
A.~Spielfiedel (2001, private communication) have shown that  for the 
isolated 
radiating atom (\mbox{ $R \rightarrow \infty$ }) this transition is not
asymptotically  forbidden as it was explicitly stated in 
Allard~et~al.~(2000). 

The line satellite shown in Fig.~\ref{beta} presents a shoulder at
 1120~\AA, a similar shape
has been obtained for the 1600~\AA\ satellite. 
In Fig.~6 of Allard~et al.~(1999) 
both theory and experiment show an oscillatory
structure between the satellite and the line, with a minimum
at about 1525~\AA.  
These oscillations are an interference 
effect (Royer~1971; Sando \& Wormhoudt~1973),
and are expected to depend on the relative velocity of the
collision and therefore on~temperature.

This satellite of Lyman~$\beta$ is quite far from the 
unperturbed Lyman~$\beta$ line center,  actually closer
to the Lyman~$\alpha$ line. It is therefore necessary to take into 
account the total
contribution of both the  Lyman~$\alpha$ and Lyman~$\beta$ wings of H
perturbed simultaneously by neutrals and protons and to study the 
variation  of this part of the Lyman series 
with the relative density of ionized and neutral atoms.

 \subsection { Lyman $\beta$ in H-H and H-H$^+$ collisions}

In Allard~et al.~(1998a) we presented Lyman~$\beta$ profiles perturbed
 by  protons. The line profile calculations were done without using
 the expansion in density and then were  valid from the center to the
 far  wing and allowed a comparison of the
amplitudes of the satellites to the line core.
 The profiles at different densities of H$^+$ were displayed in~Fig.~8.

\begin{figure}      
\hspace{-0.5cm}
\psfig{file=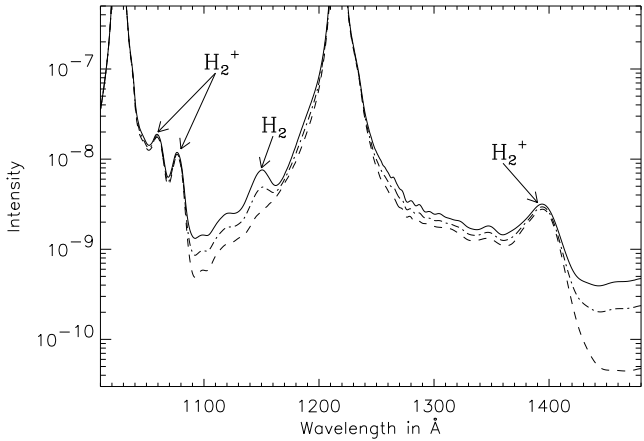,width=9.50cm}
\caption{Total profile of Lyman~$\alpha$ and 
Lyman~$\beta$ perturbed by neutral hydrogen
and protons. Three different neutral densities ($1\times10^{17}$,
$5\times10^{16}$, $1\times10^{16}$ cm$^{-3}$, top to the bottom) are
compared  for a fixed ion density ($1\times10^{16}$ cm$^{-3}$). }
\label{lytotal}
\end{figure}

In Fig.~\ref{lytotal} we show the sum of the profiles of Lyman~$\alpha$ and
Lyman~$\beta$ perturbed by collisions with neutral  hydrogen and protons  
 for different densities $n_{H}$ . 

 We can see that a ratio of 5 between the neutral and proton density is
enough to make the quasi-molecular H$_2$ satellite appear in the far wing.
Note that the 1120~\AA\ shoulder of the satellite is still visible in
the total profile.

 \section {Synthetic spectra for DA white dwarfs}

We now take into account these absorption features of the H$_2^+$ and
H$_2$ quasi-molecules to calculate synthetic spectra for DA white dwarfs.
Atmosphere models and synthetic spectra have been calculated using the 
computer programs TLUSTY and SYNSPEC (Hubeny 1988; Hubeny \& Lanz
1992, 1995). The LTE model atmospheres assume a pure hydrogen
composition including the quasi-molecular~opacities.

Below $T_{\rm eff}$=15\,000~K, the atmosphere of DA white dwarfs
become convective, and model atmospheres for these stars are usually
calculated using several variants of
the standard mixing-length theory. According 
Bergeron~et al.~(1992), we used the so-called
ML2 prescription with $\alpha=0.6$, which was shown to
provide excellent internal consistency between ultraviolet and 
optical temperatures.

\begin{figure*}      
\hspace{0.7cm}
\psfig{file=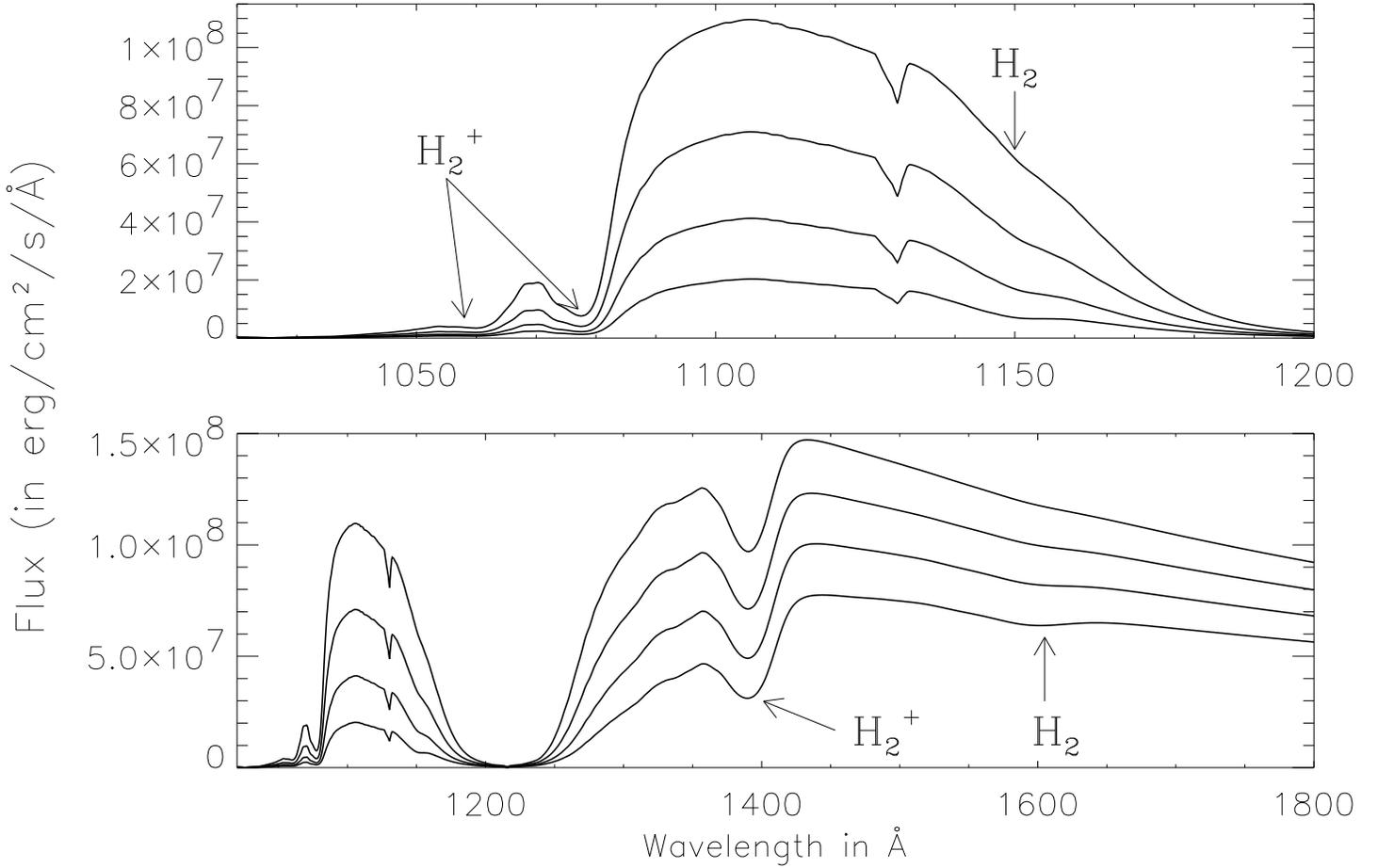,width=19.0cm}
\caption{Synthetic spectra for DA white dwarfs including
Lyman~$\alpha$  and Lyman~$\beta$ satellites, on spectral ranges 
1020~--~1200~\AA\ (upper panel) and 1020~--~1800~\AA\ (lower panel). 
$T_{\rm eff}$=14\,500, 14\,000,
13\,500, 13\,000~K (from top to bottom) and $\log g = 8.0$. 
The flux, at the stellar surface, is integrated over angles.}
\label{syntba}
\end{figure*}

The temperature range  where both
  H$_2$ and H$_2^+$  Lyman~$\beta$ satellites are visible is
  relatively small; roughly between 14500 to 11000~K
  (Fig.~\ref{syntba}-upper panel). This dependence of the theoretical
profiles on temperature  is extremely
  strong because of the relative importance of perturbations by
  neutral versus ionized hydrogen. 
 The  satellite appearance is  then very
  sensitive  to the  degree of ionization and may be used as a
  temperature  diagnostic. 

In order to see all the Lyman line satellites, synthetic
spectra  are plotted  from 1020 to 1800~\AA\ (Fig.~\ref{syntba}-lower panel).
Below $T_{\rm eff}$= 13000~K  the H$_2^+$
  starts to disappear but the broad H$_2$ satellite at 1150~\AA\ is
still present and the one at 1600~\AA\ slowly increases. 

\section {\fuse\ observations of \g231}

\subsection {Observations and data reduction}
\label{obs_and_data_red}

\g231\ (\object{WD$\,2117+539$}) is a cool DA white dwarf located at 
Galactic coordinates $l=95.0^o$ and $b=-3.3^o$; it is a relatively 
bright white dwarf: $V=12.3$. 
This target was observed with \fuse\ as part of the 
French Guaranteed Time Observing Programs (program Q210). 
Two exposures were obtained on 2001 July 3 in time-tagged photon 
address mode (TTAG) with the object in
the large aperture (LWRS). The total duration was $\sim1.6$~h
(see Table~\ref{obslog}). Details of the \fuse\ instrument may be 
found in Moos et al.~(2000) and Sahnow et al.~(2000). 

\begin{table*}
\caption[]{Observation log}
\label{obslog}
\begin{tabular}{cccccc}
\hline
Instrument & Observation & Date of observation & Aperture & 
Exposure time (sec) & Spectral extraction \\
\hline
\fuse & Q2100101001 & 2001 July  3  & LWRS ($30''\times30''$) & 2940  & 
     CalFUSE 2.0.5 \\
\fuse & Q2100101002 & 2001 July  3  & LWRS ($30''\times30''$) & 2864  & 
     CalFUSE 2.0.5 \\
\iue  &  SWP19532   & 1983 March 24 & SWLA ($10''\times20''$) & \ 720 &    
     NEWSIPS    \\
\iue  &  SWP19533   & 1983 March 24 & SWLA ($10''\times20''$) & \ 720 &    
     NEWSIPS    \\
\hline
\end{tabular}
\end{table*}

The one-dimensional spectra were extracted from the two-dimensional 
detector images and calibrated using version~2.0.5 of the CalFUSE 
pipeline. The eight \fuse\ detector segments of the two exposures (\ie\ 16 
spectra) were co-added and projected on a $0.16$~\AA-pixel base, 
\ie\ pixels about 25 times larger than the original \fuse\ detectors pixels. 
This degradation of the \fuse\ spectral resolution (typically 
$\lambda/\Delta\lambda\simeq15000$ for this kind of target in the 
large slit; see H\'ebrard et al. 2002; Wood et al. 2002) 
is of no effect on the 
shapes of the large stellar features which we study and allows us to 
increase the signal-to-noise ratio. At this resolution, no spectral 
shifts were detected between the 16 different co-added spectra. 

Poor quality edge of each segment were not included in the sum. 
We used the following spectral ranges for each segment:
$1010-1085$\AA\ (SiC1A), 
$910-992.5$~\AA\ (SiC1B), 
$922-1005$~\AA\ (SiC2A), 
$1016-1101$~\AA\ (SiC2B), 
$1004-1081.5$~\AA\ (LiF1A), 
$1096-1187$~\AA\ (LiF1B),
$1087-1181$~\AA\ (LiF2A), and 
$1012-1073$~\AA\ (LiF2B). 
As the LiF1B segment presents a known large-scale distortion 
(``the worm'', D.~Sahnow 2000, private communication) in the flux 
calibration around $1120-1170$~\AA\ (see, \eg, H\'ebrard et al.~2002), 
we normalized the large-scale shape of the LiF1B segment to the shape 
of the LiF2A segment, using a polynomial fit. 
We also did not include the part of the LiF2A spectra in the 
ranges $1134.6-1135.6$~\AA\ and $1151.5-1153$~\AA\ because they were 
deteriorated by the so-called ``walk'' problem, which is a distortion 
in the \fuse\ spectra at airglow wavelengths ($\lambda\;1134$~\AA\ 
\ion{N}{i} and $\lambda\;1152$~\AA\ \ion{O}{i} lines in the present case)
caused by variation of position with pulse height (D.~Sahnow 2001, 
private communication).
The final \fuse\ spectrum is plotted in Fig.~\ref{fuse}. Note that the 
interstellar \ion{N}{i} $\lambda\;1134$~\AA\ triplet is detected on this 
line of sight.

Since 2001, repeated observations of standard white dwarfs indicate 
a slow degradation in the effective area of the \fuse\ spectrograph. The 
decline affects the flux up to 20\%, 
and that mainly for the LiF2A channel. A dime-dependent flux 
correction is in development for a future version of the pipeline, but is 
not included in the CalFUSE version 2.0.5.

In order to complement the $910-1187$~\AA\ \fuse\ spectrum, we retrieved 
two \iue\ low dispersion spectra of \g231\ from the 
STScI MAST archive. These spectra were obtained in March 1983 
and they were reduced using the NEWSIPS processing (see Table~\ref{obslog}). 
They yield the spectral coverage $1150-1980$~\AA\ at a resolution of 
about $7$~\AA\ (Wegner 1984).
The flux calibration of the \fuse\ and the \iue\ spectra are in 
good agreement (see Fig.~\ref{fuse}, lower~panel).

\subsection {Comparison with theoretical spectra}

As a change in temperature can be compensated by a change in
 surface gravity  we have used $\log g$ determined from optical analysis.
Holberg~et al.~(1998) give (14\,490; 7.85) for ($T_{\rm eff}$; $\log g$),
the atmospheric parameters being determined from
 spectroscopic observations by Bergeron~et al.~(1992). 
Model atmospheres and synthetic spectra were calculated in the  range
 15\,500  to 14\,500~K by steps of 100K,  using  $\log g$=7.85.

\begin{figure*}      
\psfig{file=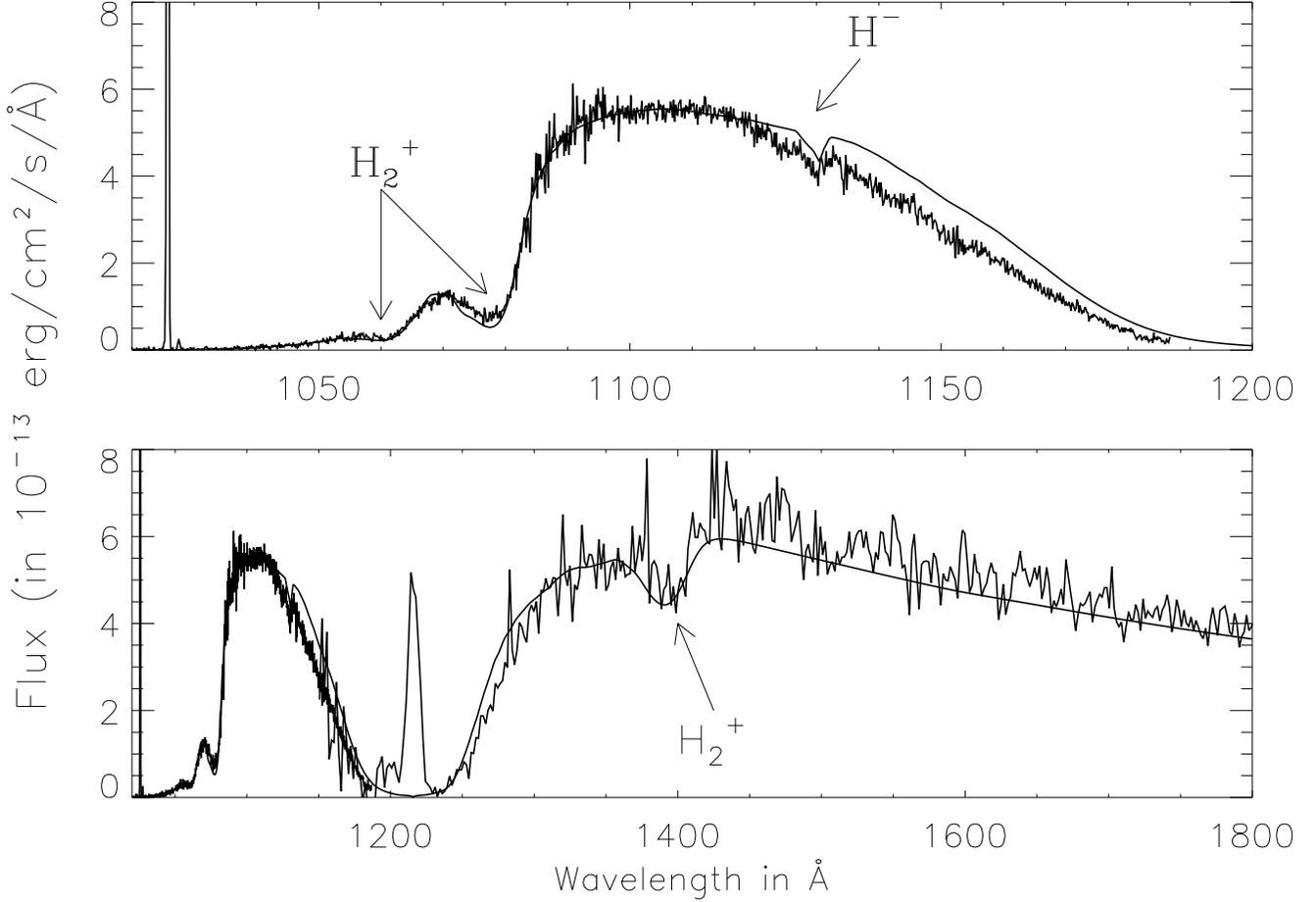,width=19.0cm}
\caption{\fuse\ (1020~--~1187~\AA, upper and lower panels) and \iue\ 
(1150~--~1800~\AA, lower panel)
spectra of \g231\ compared with a theoretical model  
for  $T_{\rm eff}$= 14\,800~K  and 
$\log g = 7.85$. Emissions at Lyman~$\alpha$, Lyman~$\beta$, and 1027.5~\AA\ 
are due to \ion{H}{i} and \ion{O}{i} airglow. The difference between the data 
and the model between 1120 and 1180~\AA\ is mainly due to an uncorrected 
decreasing of sensitivity on the \fuse\ segment LiF2A 
(see Sect.~\ref{obs_and_data_red}). 
Notice the H$^-$ feature at 1130~\AA.}
\label{fuse}
\end{figure*}
H$^-$ opacity (Wishart~1979) has been added in the atmosphere model
calculation and gives a
feature at 1130~\AA\ also present in the \fuse\ observation 
(see Fig.~\ref{fuse}). 

Fig.~\ref{fuse}-upper panel shows the comparison of the \fuse\
observation with our best fit obtained for $T_{\rm eff}$=14\,800~K. 
The decreasing of sensitivity on the segment LiF2A 
(Sect.~\ref{obs_and_data_red}) is the main explanation of the difference 
between the data and the model between 1120 and 1180~\AA. 
The H$_2$ satellite (1150~\AA) is no more visible, as expected, in this
range of temperature.

Fig.~\ref{fuse} shows a good agreement between the \fuse\
and  \iue\ observations of Lyman lines and predicted spectra over the
whole range from 1020 to 1800~\AA\ using atmospheric
 parameters determined from the optical range. 

The feature near $\lambda\,995\,$\AA\ due to a Lyman~$\gamma$ 
satellite, which is visible in the spectra of hotter objects, 
is no more present in the temperature range of \g231\ 
(the flux of \g231\ is below 
$1\times10^{-14}\,$erg~cm$^{-2}$~s$^{-1}$~\AA$^{-1}$
at $995\,$\AA). 
It was visible in \fuse\ observations of 
\object{CD\,$-$38$^{\circ}$\,10980} (Wolff~et al.~2001), {\it HUT} 
spectrum of \object{Wolf$\;1346$} (Koester~et al.~1996), 
and in some {\it ORFEUS} spectra (Koester~et al.~1998). 
It was also recently seen in \fuse\ spectra of \object{Sirius~B} 
(Holberg et al. 2002). 
This feature, due to a Lyman~$\gamma$ satellite,  is 
now included in our calculations. We have shown that  a larger
temperature is required to get it detectable (Allard~et al.~2002).  

\section {Conclusion}

From theoretical profiles including the new Lyman~$\beta$ opacities it
can be predicted that  Lyman~$\beta$ satellites should be detectable
roughly between 25000~K to 11000~K. Within this range in a very small 
domain of temperatures, the whole profile should present H$_2$ and
H$_2^+$ satellites of the Lyman~$\alpha$ and Lyman~$\beta$ lines.
\fuse\ observations of cooler white dwarfs in the  ZZ$\,$Ceti  range 
 stars would offer the best opportunity to determine accurate 
stellar parameters as $T_{\rm eff}$ and  $\log g$ for these~stars.

\begin{acknowledgements}
This work is based on data obtained for the French Guaranteed Time
by the NASA-CNES-CSA \fuse\ mission operated by the Johns Hopkins University.
I. H. would like to thank the PNPS (Programme National de Physique Stellaire)
for travel support to IAP where part of this work was done. 
We would like to thank Jean Dupuis for useful comments on this work. 
\end{acknowledgements}


\begin{thebibliography}{}


  \bibitem{} Allard, N. F., \& Kielkopf,~J. 1982, Rev.~Mod.~Phys., 54, 1103

  \bibitem{} Allard, N. F., Koester, D., Feautrier, N., \& Spielfiedel, A.
     1994, A\&A, 200, 58

  \bibitem{} Allard, N. F., Kielkopf, J. F., \& Feautrier, N.
     1998a, A\&A, 330, 782 

  \bibitem{} Allard, N. F., Drira, I., Gerbaldi, M., Kielkopf, J. F., \&
  Spielfiedel, A. 1998b, A\&A, 335, 1124

  \bibitem{} Allard, N. F., Royer, A., Kielkopf, J. F., \& Feautrier, N.
     1999, Phys. Rev. A, 60, 1021

  \bibitem{} Allard,~N. F., Kielkopf,~J. F., Drira,~I., \& Schmelcher,~P. 
    2000, Eur. Phys. J. D, 12, 263 

 \bibitem{} Allard,~N. F., Bourdreux,~S., Kielkopf,~J. F., et al. 
2002, 13th European Workshop on White Dwarfs, Napoli, 24-28 juin 2002 

 \bibitem{} Bergeron, P., Saffer, R., \& Liebert,~J. 1992, ApJ, 394, 228 

  \bibitem{} Bergeron, P., Wesemael, F., Lamontagne, R., et al. 
  1995, ApJ, 449, 258 

   \bibitem{} Detmer, T., Schmelcher, P., \& Cederbaum, L. S. 1998,
    J. Chem. Phys., 109, 9694  

    \bibitem{} Drira, I. 1999, J. Mol. Spectroscopy, 198, 52

     \bibitem{} H\'ebrard, G. et al. 2002, ApJS 140, 103

  \bibitem{} Holberg, J. B., Barstow,~M. A., \& Sion,~E. M., 1998, 
ApJS, 119, 207

  \bibitem{} Holberg, J. B., Kruk, J. W., Barstow, M. A., et al. 
  2002, \fuse\ Science and Data Workshop, 
  JHU, Baltimore, p. 56

     \bibitem{} Hubeny, I. 1988, Comp. Phys. Comm., 52, 103

     \bibitem{} Hubeny I., \& Lanz, T. 1992, A\&A, 262, 501

     \bibitem{} Hubeny I., \& Lanz, T. 1995, ApJ, 439, 875

  \bibitem{} Kielkopf, J. F., \& Allard, N. F. 1995, ApJ, 450, L75

  \bibitem{} Kielkopf, J. F., \& Allard, N. F. 1998,  Phys. Rev. A, 58, 4416

  \bibitem{}  Koester, D., \& Allard, N. F. 1993, in {\it White Dwarfs: 
     Advances in Observation and Theory}, ed. M. Barstow (Kluwer: Dordrecht), 
     p. 237

  \bibitem{} Koester, D., Allard, N. F., \& Vauclair, G. 1994, A\&A, 291, L9

  \bibitem{} Koester, D., Finley, D.~S., Allard, N. F., Kruk, J. W., 
     \& Kimble, R. A. 1996, ApJ, 463, L93

   \bibitem{} Koester,~D., Sperhake,~U., Allard,~N. F., Finley,~D. S.,
      \& Jordan,~S. 1998, A\&A, 336, 276 

  \bibitem{} Madsen, M. M., \& Peek, J. M. 1971, Atomic Data, 2, 171

  \bibitem{} Moos, H. W. et al. 2000, ApJ, 538, L1

  \bibitem{} Ramaker, D. E., \& Peek, J. M. 1972, J. Phys. B, 5, 2175

  \bibitem{}  Royer, A. 1971, Phys. Rev. A, 43, 499

  \bibitem{} Sahnow, D.\ J.\ et al.\ 2000, ApJ, 538, L7 

  \bibitem{} Sando,~K.~M., \& Wormhoudt,~J.~G. 1973, Phys. Rev. A, 7, 1889 

 \bibitem{}  Wishart, A. W. 1979, MNRAS 187, 59p

  \bibitem{} Wegner, G. 1984, AJ, 89, 1050

  \bibitem{} Wolff,~B., Kruk,~J. W., Koester,~D., et al. 2001, A\&A, 373, 674

  \bibitem{} Wood, B. E., Linsky, J. L., H\'ebrard, G., et al. 
  2002, ApJS 140, 91

\end{thebibliography}
\end{document}